\documentclass[prd,twocolumn,showpacs,showkeys,preprintnumbers,superscriptaddress,longbibliography]{revtex4-2}

\usepackage{soul}
\usepackage{amsfonts} 
\usepackage{amssymb} 
\usepackage{amsmath} 
\usepackage{graphicx} 
\usepackage{subfigure} 
\usepackage{array} 
\usepackage{dcolumn} 
\usepackage{bm} 
\usepackage{latexsym} 
\usepackage{longtable} 
\usepackage{hyperref} 
\usepackage{bbold}
\usepackage{color}

\usepackage{comment}
\usepackage{mathtools}
\usepackage{algorithm}
\usepackage[noend]{algpseudocode}
\usepackage{lipsum}

\newcommand{\GF}{G_\mathrm{F}}

\usepackage[dvipsnames,usenames]{xcolor}
\input{Qcircuit}



\begin{document}

\title{Simulation of Collective Neutrino Oscillations on a Quantum Computer}

\author{Benjamin Hall}
\affiliation{Facility for Rare Isotope Beams (FRIB), Michigan State University, East Lansing, MI 48824, USA}
\author{Alessandro Roggero}
\affiliation{Institute for Nuclear Theory, University of Washington, Seattle, WA 98195, USA}
\affiliation{InQubator for Quantum Simulation (IQuS), Department of Physics, University of Washington, Seattle, WA 98195, USA}
\author{Alessandro Baroni}
\affiliation{Theoretical Division, Los Alamos National Laboratory, Los Alamos, NM 87545, USA}
\author{Joseph Carlson}
\affiliation{Theoretical Division, Los Alamos National Laboratory, Los Alamos, NM 87545, USA}

\preprint{IQuS@UW-21-03,LA-UR-21-20599,INT-PUB-21-005}

\date{\today}

\begin{abstract}
In astrophysical scenarios with large neutrino density, like supernovae and the early universe, the presence of neutrino-neutrino interactions can give rise to collective flavor oscillations in the out-of-equilibrium collective dynamics of a neutrino cloud. The role of quantum correlations in these phenomena is not yet well understood, in large part due to complications in solving for the real-time evolution of the strongly coupled many-body system.
Future fault-tolerant quantum computers hold the promise to overcome much of these limitations and provide direct access to the correlated neutrino dynamic. In this work, we present the first simulation of a small system of interacting neutrinos using current generation quantum devices. We introduce a strategy to overcome limitations in the natural connectivity of the qubits and use it to track the evolution of entanglement in real-time. The results show the critical importance of error-mitigation techniques to extract meaningful results for entanglement measures using noisy, near term, quantum devices.
\end{abstract}

\maketitle

\section{Introduction}

The flavor evolution of neutrinos in dense astrophysical environments have, by now, a long history. It has been pointed out by Pantelone, Raffelt, and Sigl~\cite{Pantaleone:1992eq,Sigl:1993} and others that, through forward scattering, neutrinos can exchange their flavors. Given an anisotropic initial distribution in energy and/or angle as found in supernovae, neutron star mergers, or the early universe, the neutrino energy flux versus energy and flavor may be impacted by this non-trivial quantum many-body evolution.
This can in turn affect the dynamics of these environments and other observable signatures, including nucleosynthesis in the ejected material (see~\cite{Duan:2010bg,Chakraborty2016b} for recent reviews).

The Hamiltonian for neutrino flavor evolution in a dense neutrino environment includes three terms: the vacuum mixing that has been determined from solar and accelerator neutrino experiments~\cite{GonzalezGarcia:2003}, the forward scattering in matter leading to the well known MSW effect~\cite{Wolfenstein:1977ue,Mikheev:1986gs},
and neutrino-neutrino forward scattering.

In the neutrino flavor basis, the vacuum term includes diagonal contributions describing the mass differences between different neutrino flavors and an off-diagonal term characterized by a mixing angle $\theta_v$.
The interaction describing forward scattering in matter is diagonal in the flavor basis.
The neutrino-neutrino interaction can exchange
flavors of two neutrinos and has a forward scattering amplitude that depends on the angle between their momenta. 
For the two-flavor case considered here, this interaction is proportional to the dot product $\vec{\sigma}_i \cdot \vec{\sigma}_j$ of the SU(2) matrices
describing the different flavor amplitudes of the two neutrinos
\begin{equation}
\label{eq:fws_int}
V_{ij}\propto \left(1-\frac{\vec{q}_i\cdot\vec{q}_j}{\|\vec{q}_i\|\|\vec{q}_j\|}\right)\vec{\sigma}_i\cdot\vec{\sigma}_j\;.
\end{equation}
Here we denoted by $\vec{q}_k$ the momentum of the $k$-th neutrino and with  $\vec{\sigma}_k=(\sigma_k^x,\sigma_k^y,\sigma_k^z)$ the vector of Pauli operators acting on its amplitude. Generalization to the three-flavor case is straightforward in principle;
here we assume the $\mu$ and $\tau$ flavors evolve similarly.

For this simplified two-flavor case, we seek to understand the time and space evolution of the set of amplitudes from a Schr\"{o}dinger equation:
\begin{equation}
\label{eq:schroedinger-eq}
\ket{\Phi (t)} = \exp [ -iH t] \ket{\Phi_0},
\end{equation}
with $H$ the total Hamiltonian including both the vacuum and forward-scattering interaction contributions. For simplicity here we 
consider $\ket{\Phi_0}$ to be a product state, but generalizations to arbitrary states are straightforward.

Most often these quantum equations have been treated on the mean-field level by replacing one of the spin operators in Eq.~\eqref{eq:fws_int} by its expectation value, yielding a set of non-linear coupled differential equations. This makes the calculations tractable for several hundred energies and angles on modern computers (see eg.~\cite{Duan:2006}).
More recently, studies of neutrino propagation as a quantum many-body problem have appeared,
including for example \cite{Bell2003,Friedland2003,sawyer2004classical,Pehlivan2011,Rrapaj2020,Cervia:2019,Roggero2021a,Roggero2021b}. These works highlight the importance of understanding the role of quantum correlations, such as entanglement, in order to quantify beyond mean-field effects in out-of-equilibrium neutrino simulations. A direct solution of the Schr\"{o}dinger equation in Eq.~\eqref{eq:schroedinger-eq}, for a system of $N$ configurations in energy and angle, incurs a computational cost that is exponential in $N$. This has limited early explorations of the problem to systems with $N=\mathcal{O}(10)$ neutrinos. An alternative to reach larger system sizes, explored recently by one of us in Refs.~\cite{Roggero2021a,Roggero2021b}, employs a Matrix Product State representation for $\ket{\Phi(t)}$ which allows one to track the exact time evolution in situations where entanglement never grows too much. For conditions leading to strong entanglement instead, simulations on digital/analog quantum computers have the potential to tackle the full neutrino dynamics while still enjoying a polynomial computational cost in system size $N$~\cite{Lloyd96}.

In this work we explore the time-dependent many-body evolution of the neutrinos on a current-generation digital quantum computer. In Sec.~\ref{sec:spin_model} we introduce in more detail the SU(2) spin model used to describe collective neutrino oscillations and describe an implementation of the time evolution operator appearing in Eq.~\eqref{eq:schroedinger-eq} suitable for an array of qubits with linear connectivity. We present the results obtained for a a small system with $N=4$ neutrino amplitudes in Sec.~\ref{sec:results} and provide a summary and conclusions in Sec.~\ref{sec:conclusion}.

\section{Spin model for neutrino oscillations}
\label{sec:spin_model}

For the simplified two-flavor case studied here, the state of the system can be described as an amplitude for a neutrino of each energy $E_i$ (equal to the magnitude of momentum $\|\vec{q}_i\|$) and direction of momentum (denoted by $\hat{q}_i$), with $\alpha_\uparrow$ and $\alpha_\downarrow$ describing the amplitude of being in the electron flavor or in a heavy ($\mu$ or $\tau$) flavor respectively. These two amplitudes can be encoded in an SU(2) spinor
basis. In this basis, the Hamiltonian can be written in terms of Pauli operators as the sum of a one-body term, describing both vacuum oscillations and forward scattering in matter, 
\begin{equation}
\label{eq:Ham_init}
H_1  = \frac{1}{2} \sum_i \left[ \left(-\Delta_i\cos{2 \theta_v} + A\right)\sigma^z_i + \Delta_i\sin {2 \theta_v} \sigma^x_i \right]\;,
\end{equation}
and a two-body term, coming from the neutrino-neutrino forward-scattering potential $V_{ij}$ from Eq.~\eqref{eq:fws_int}, which takes the following form~\cite{Pehlivan2011}
\begin{equation}
\begin{split}
H_2 & =  \sum_{i<j} \eta  [1- \hat{q}_i \cdot \hat{q}_j ] \vec{\sigma}_i \cdot \vec{\sigma}_j\;.
\end{split}
\end{equation}
In the one-body term, $\theta_v$ represents the vacuum mixing angle, while the strength is given by $\Delta_i = \delta m^2/(2 E_i)$ with $\delta m^2$ the mass squared difference for neutrinos of different flavor. The matter potential enters as the diagonal contribution in the one-body term through the constant $A = \sqrt{2} \GF n_e$, with $\GF$ the Fermi coupling constant and $n_e$ the electron density.

As described in the introduction, the two-body term is a sum over spin-spin interactions with a coupling depending upon the relative angle between them. The overall strength depends on the neutrino density as
\begin{equation}
\label{eq:eta}
\eta=\frac{\GF }{\sqrt{2}V}=\frac{\GF n_\nu}{\sqrt{2}N}\;,
\end{equation}
with $N$ the number of neutrino momenta considered, given by the neutrino density $n_\nu$ times the quantization volume $V$. The Hamiltonian is similar to a Heisenberg model, but the two-body term
is all-to-all rather than nearest neighbor. Its coupling strength $\eta\propto1/N$ assures that the energy of the system is extensive. This allows us to obtain a well-defined many-body solution, in the limit of large
numbers of neutrino momenta by extrapolating in system size $N$.

Currently available quantum devices are able to perform only a relatively limited number of operations while maintaining a high fidelity~\cite{Preskill2018}, this in turn poses limits on the maximum time that could be reached in the simulation of neutrino dynamics. Given this practical constraint, it is then useful to consider a test case where the one- and two-body interaction terms are similar in magnitude and the evolution can occur rapidly. 
An example is the environment of order $\approx100$ km from the surface of a proto-neutron star in a core collapse supernovae.  Here the background matter density has decreased to a point where its contribution to the Hamiltonian is similar in magnitude to the neutrino-neutrino forward scattering.  The relative angles of neutrino propagation are fairly small as neutrinos are emitted from a typical proto-neutron star radius of order 10 km. In the neutrino bulb model~\cite{Duan:2006} one further assumes the evolution in a supernovae depends only on the energy and the angle from the normal. Averaging over the azimuthal angles results in an average
coupling $\langle 1-{\hat {q}}_i \cdot {\hat {q}}_j \rangle = 1 - \cos(\theta_i) \cos(\theta_j).$
Often a further simplification, usually called single-angle approximation, is made where an average coupling is taken between all pairs of neutrinos, resulting in a two-body term simply related to the square of the total spin $S$ of the many-body state.

For our test case, we take a monochromatic neutrino beam with energy $E_\nu=\delta m^2/(4\eta)$ and measure energies in units of the two-body coupling $\eta$. In order to avoid the symmetries introduced by the single angle approximation, we employ an anisotropic distribution of momentum directions
using a simple grid of angles with
\begin{equation}
\theta_{pq} = \arccos(0.9) \frac{|p-q|}{N-1}\;.
\end{equation}
This is similar to the standard bulb model as the relative couplings $1- \cos(\theta_{pq})$ are small.

The final Hamiltonian for the simple model we implement here can be written compactly, in units of $\eta$, as
\begin{equation}
\label{eq:spin_ham}
H = \sum_{k=1}^N \vec{b} \cdot \vec{\sigma}_k + \sum_{p<q}^N J_{pq} \vec{\sigma}_p\cdot\vec{\sigma}_q\;,
\end{equation}
with the external field $\vec{b}=\left(\sqrt{1-0.925^2},0,-0.925\right)$ obtained by choosing the mixing angle $\theta_v=0.195$ and pair coupling matrix $J_{pq}=\left(1-\cos\left(\theta_{pq}\right)\right)$. Note that in this model we set the matter potential $A$ in the one-body contribution to the Hamiltonian Eq.~\eqref{eq:Ham_init} to zero.

\subsection{Real time evolution}
\label{sec:real_time}

The major challenge in implementing the time evolution in Eq.~\eqref{eq:schroedinger-eq} in a quantum simulation is to find an accurate approximation to the evolution operator $U(t)=\exp[-iHt]$ that can also be decomposed efficiently into local unitary operations~\cite{Lloyd96}. A simple and popular approach is to use a first-order Trotter-Suzuki decomposition~\cite{Suzuki91} of the propagator leading to the approximation
\begin{equation}
\label{eq:trotter1}
U_1(t) = 
\prod_{j=1}^Ne^{-it\vec{b}\cdot\vec{\sigma}_j}
\prod_{p<q}^N e^{-itJ_{pq}\vec{\sigma}_p\cdot\vec{\sigma}_q}\;,
\end{equation}
which is correct up to an additive error $\epsilon=\mathcal{O}(t^2)$. Past experience with the Euclidean version of this evolution operator in Quantum Monte Carlo suggests that a better approximation to the full propagator $U(t)$ can be obtained by using the exact propagators for pairs (see eg.~\cite{Ceperley1995,Carlson2015}). In order to construct this alternative approximation, we first rewrite the Hamiltonian in Eq.~\eqref{eq:spin_ham} manifestly as a sum of $\binom{N}{2}$ two-body Hamiltonians acting on each pair of qubits
\begin{equation}
\label{eq:ham_decomp}
H =\sum_{p<q}^N \left(\frac{\vec{b}\cdot\left(\vec{\sigma}_p+\vec{\sigma}_q\right)}{N-1} + J_{pq}\;\vec{\sigma}_k\cdot\vec{\sigma}_q\right) \coloneqq \sum_{p<q}^N h_{pq}\;.
\end{equation}
We can then define an approximate propagator $U_2$ using the exact pair propagator as follows
\begin{equation}
\label{eq:pair_prop_app}
U_2(t) = \prod_{p<q}^N e^{-ith_{pq}} \coloneqq \prod_{p<q}^N u_{pq}\;.
\end{equation}
Note that the implementation of this operator is efficient since each pair Hamiltonian acts non-trivially only on a $4\times4$ subset of the total Hilbert space and therefore, as shown for instance in Refs.~\cite{vidal2004,vatan2004}, can be implemented exactly using at most $3$ entangling operations. Note that the error in this approximation still scales as $\mathcal{O}(t^2)$ but now with a possibly reduced prefactor. In Appendix~\ref{app:pair_prop} we present a direct comparison between the two approximations. Finally, the approximation order could also be improved by symmetrizing over the ordering of operators or by applying symmetry transformations (see eg.~\cite{Tran2021}).

Owing to the long range of the interactions, a naive implementation of this scheme will require either a device with all-to-all connectivity (like trapped ion systems~\cite{Monroe2019}) or an extensive use of the SWAP operation, represented in matrix form as
\begin{equation}
\text{SWAP} = \begin{pmatrix}
1&0&0&0\\
0&0&1&0\\
0&1&0&0\\
0&0&0&1\\
\end{pmatrix}\;.
\end{equation}
The effect of this operation is to exchange the state of two qubits. One can then use this operation to bring a pair of qubits that we want to interact close to each other by applying a sequence of SWAP gates of order $N$. Since we need to apply all possible pair interactions, we will show that it is actually possible to carry out a complete step, under the unitary in Eq.~\eqref{eq:pair_prop_app}, without incurring any overhead due to the application of the SWAP operations. The scheme is inspired by the more general fermionic swap network construction presented in Ref.~\cite{Kivlichan2018}. 

\begin{figure}
 \centering
 \includegraphics[width=0.45\textwidth]{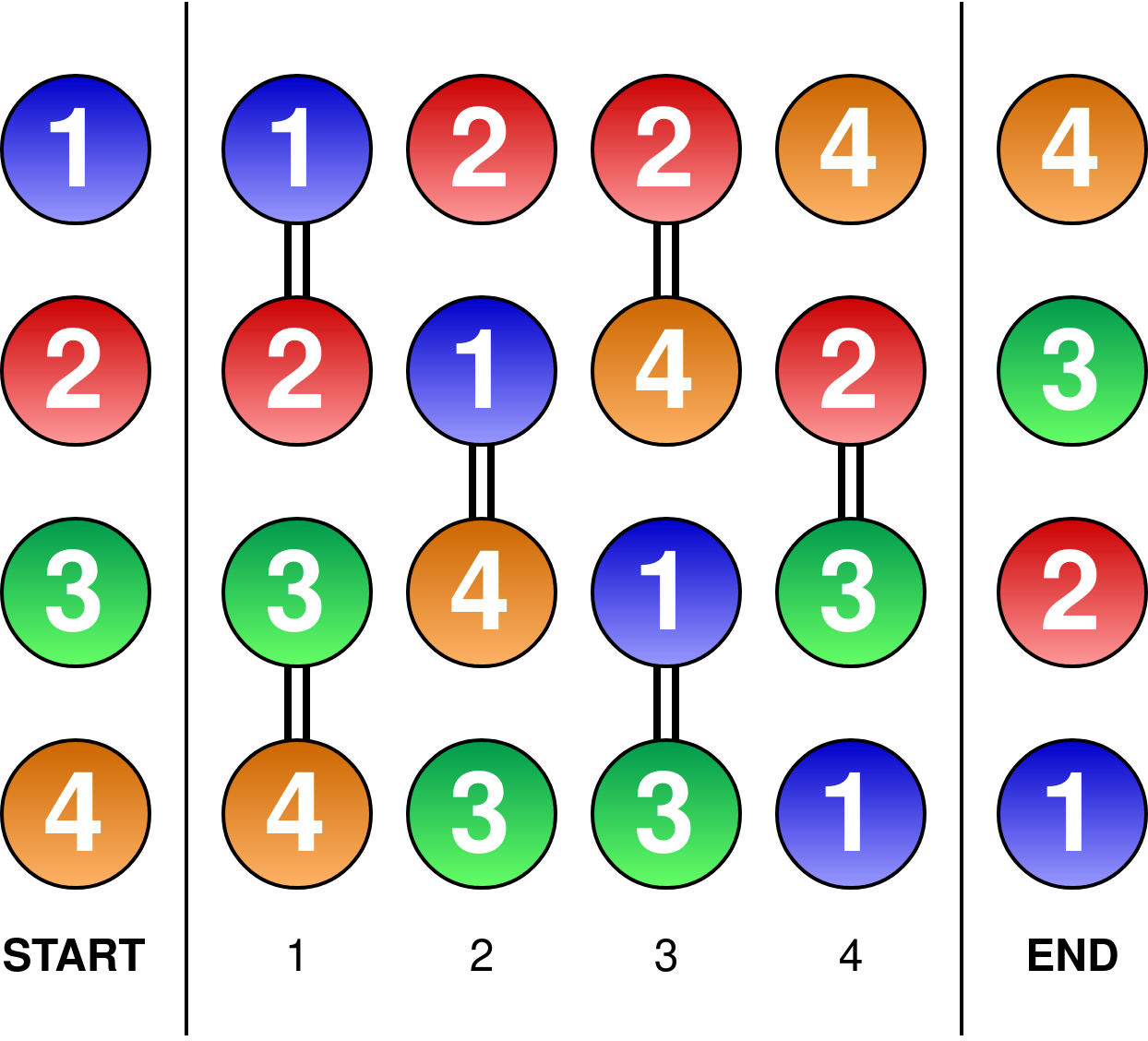}
 \caption{(Color online) Pictorial representation of the swap network used in our simulation in the case of $N=4$ neutrinos.}
\label{fig:swap_network}
\end{figure}

We illustrate this idea using the diagram shown in Fig.~\ref{fig:swap_network} for a simple case with $N=4$ neutrinos. Starting from the initial state on the left, we first apply the unitaries $u_{pq}$ from Eq.~\eqref{eq:pair_prop_app} to the odd bonds: for the $N=4$ case, these are the bonds between the $(1,2)$ and $(3,4)$ pairs of qubits. Before moving to the next pairs, we also apply a SWAP operation to the same pairs we just acted upon. The resulting unitary operation is denoted as a double line joining qubits in Fig.~\ref{fig:swap_network} and the net effect is that at the next step the qubits that have interacted get interchanged. Given the discussion following Eq.~\eqref{eq:pair_prop_app} above, this modified two-qubit unitary still requires at most 3 entangling operations.
At the end of a sequence of $N$ such combined operations we will have implemented the full unitary in Eq.~\eqref{eq:pair_prop_app} while, at the same, we inverted the ordering of qubits, as shown in Fig.~\ref{fig:swap_network}. This approach requires exactly the minimum number $\binom{N}{2}$ of nearest-neighbor pair operations, while the shifted ordering can be controlled completely, and efficiently, by classical means. Note that if we were to repeat at this point the same swap network in reverse order, the full unitary will correspond to a second order step for time $2t$ and the final ordering of qubits will be restored to it's original one. This is the strategy used in Refs.~\cite{Roggero2021a,Roggero2021b} to study the neutrino Hamiltonian with Matrix Product States. In this first implementation on quantum hardware, we focus instead on a single, linear-order, time step.

Note that since we are only using nearest neighbor two-qubit gates, the total number of entangling gates required for a full time evolution step is bounded from above by $3\binom{N}{2}$ while the maximum number of single qubit operations is bounded by $15\binom{N}{2}$. As we will see in the results presented below, the presence of a large number of arbitrary single qubit rotations seems to be the limiting factor in implementing this scheme on the quantum device we used in this first exploration.

\begin{figure}
    \centering
    \includegraphics[width=0.25\textwidth]{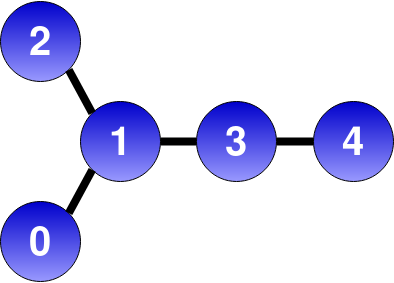}
    \caption{(Color online) Layout of the IBM Quantum Canary Processor Vigo~\cite{IBMQ_Vigo}. Shown are the five qubits, labeled from 0 to 4, and their connectivity denoted as solid black lines.}
    \label{fig:vigo}
\end{figure}

\section{Results with four neutrinos}
\label{sec:results}

In order to study the build up of correlations and entanglement generated by the time-evolution under the Hamiltonian in Eq.~\eqref{eq:spin_ham}, we first initialize a system of $N=4$ qubits in the following product state 
\begin{equation}
\ket{\Phi_0} = \ket{e}\otimes\ket{e}\otimes\ket{x}\otimes\ket{x} = \ket{\uparrow\uparrow\downarrow\downarrow}\;.
\end{equation}
We then preform one step of time evolution for time $t$ by applying the $N$ layers of nearest-neighbor gates as described in the previous section. This corresponds to a single Trotter-Suzuki step for different values of the time-step.
The four SU(2) spins representing the neutrinos are mapped to qubits $(2,1,3,4)$ on the IBMQ Vigo quantum processor~\cite{IBMQ_Vigo}, whose connectivity is schematically depicted in Fig.~\ref{fig:vigo}. The resulting qubits are linearly connected, allowing us to carry out natively the complete simulation scheme depicted in Fig.~\ref{fig:swap_network} above.

\begin{figure}[t]
 \centering
 \includegraphics[width=0.49\textwidth]{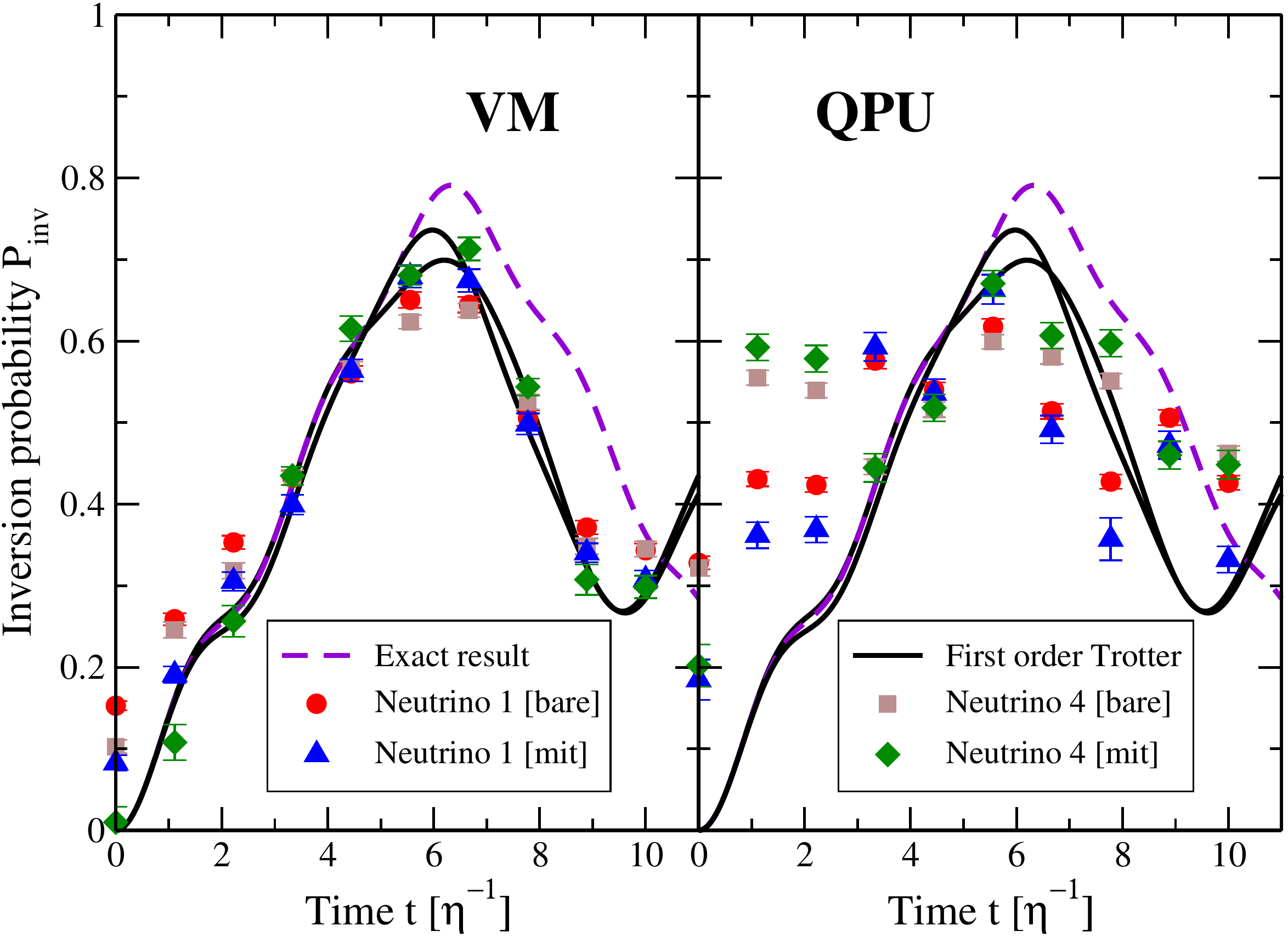}
 \caption{(Color online) Inversion probability $P_{inv(t)}$ for neutrinos 1 and 4: the red circle and brown square correspond to the bare results, the blue triangle and the green diamond are obtained after error mitigation(see text). The left panel (VM) are virtual machine results while the right panel (QPU) are results obtained on the Vigo~\cite{IBMQ_Vigo} quantum device.}
\label{fig:pop_03}
\end{figure}

\begin{figure}[t]
 \centering
 \includegraphics[width=0.49\textwidth]{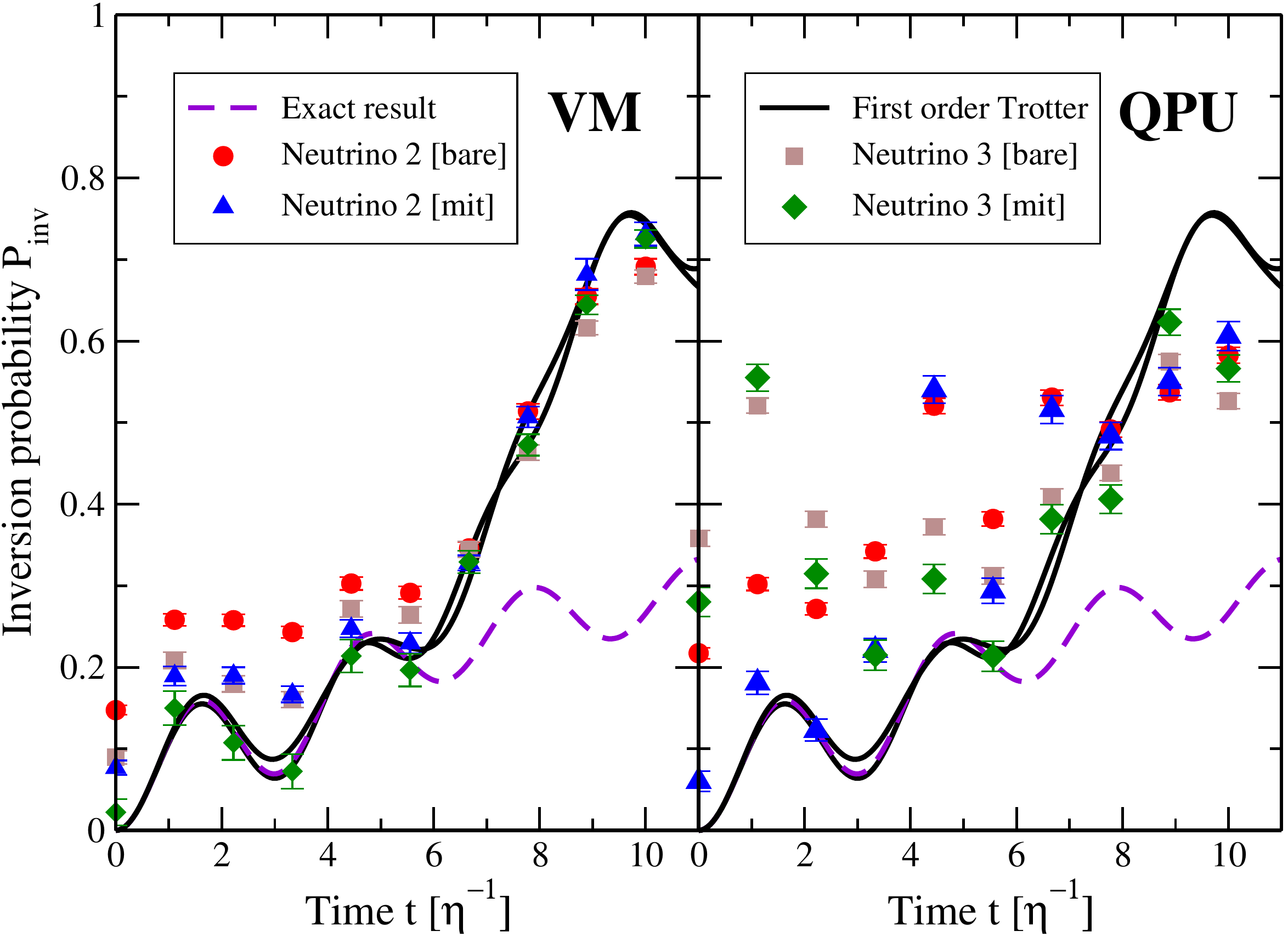}
 \caption{(Color online) Inversion probability $P_{inv}(t)$ for neutrinos 2 and 3. The notation is the same as for Fig.~\ref{fig:pop_03}.}
\label{fig:pop_12}
\end{figure}

The first observable we compute is the flavor polarization of individual neutrinos as a function of time. Since the spin Hamiltonian in Eq.~\eqref{eq:spin_ham} is invariant under the simultaneous exchanges $1\leftrightarrow4$ and $2\leftrightarrow3$, while the flavor content of the initial state $\ket{\Phi_0}$ gets reversed by it, we show directly the probability $P_{\text{inv}}(t)$ to find a neutrino in the opposite flavor to the starting one it had at $t=0$. In the limit of no error, $P_{\text{inv}}(t)$ should then by the same for the pair of neutrinos $(1,4)$ and $(2,3)$. The errors in the approximation of the propagator in Eq.~\eqref{eq:pair_prop_app} do not exactly follow this symmetry, with deviations in the range $3-7\%$. We show the results for $P_{inv}$ obtained with the approximate evolution operator $U_2(t)$ as solid black lines in Fig.~\ref{fig:pop_03}, for the pair $(1,4)$, and in Fig.~\ref{fig:pop_12} for the pair $(2,3)$. The ideal, and symmetric, result is shown instead as a purple dashed line. We see that the approximation error is very small up to relatively large time $\eta t\approx 6$. As we discuss more in detail in Appendix~\ref{app:pair_prop}, this is in large part an effect of using the pair propagator $U_2(t)$ instead of the naive first order formula in Eq.~\eqref{eq:trotter1}. 

The results shown in Fig.~\ref{fig:pop_03} and Fig.~\ref{fig:pop_12} were obtained using either the real quantum device (right panels denoted QPU) or a local virtual machine simulation employing the noise model implemented in Qiskit~\cite{qiskit} (left panels denoted by VM) initialized with calibration data from the device. In both plots we report the results (denoted by [bare]) obtained directly from the simulation and including only statistical errors coming from a finite sample size (here and in the rest of the paper we use $8192$ repetition, or ``shots", for every data point), as well as results obtained after performing error mitigation (denoted by [mit]). This corresponds to a final post-processing step that attempts to reduce the influence of the two main sources of errors: the read-out errors associated with the imperfect measurement apparatus and the gate error associated with the application of entangling gates. The latter error is dealt with using a zero noise extrapolation strategy (see~\cite{Endo2018,Dumitrescu2018} and Appendix~\ref{app:error_mit} for additional details). 

As seen also in previous similar calculations (see for instance~\cite{roggero2020A,Roggero_nptodg}), the VM results obtained using the simulated noise are much closer to the ideal result than those obtained with the real device. This is also reflected in the fact that the error mitigation protocol is not as successful with the real QPU data as it is with the simulated VM data. This behaviour is possibly linked to the substantial noise caused by the presence of a large number of single qubit operations (up to 90 rotations for time evolution + 2 for state preparation) together with the relatively large CNOT count of 18. In fact, the performance of error mitigation for the results with the largest state preparation circuits presented in~\cite{Roggero_nptodg} is superior to the one obtained here, despite the use of the same device, the same error mitigation strategy and a comparable number of entangling gates (15 CNOT in that case) while the number of rotations was only 14. This suggests coherent errors constitute a considerable fraction of the overall error seen in the results above.

\begin{figure}[t]
 \centering
 \includegraphics[width=0.49\textwidth]{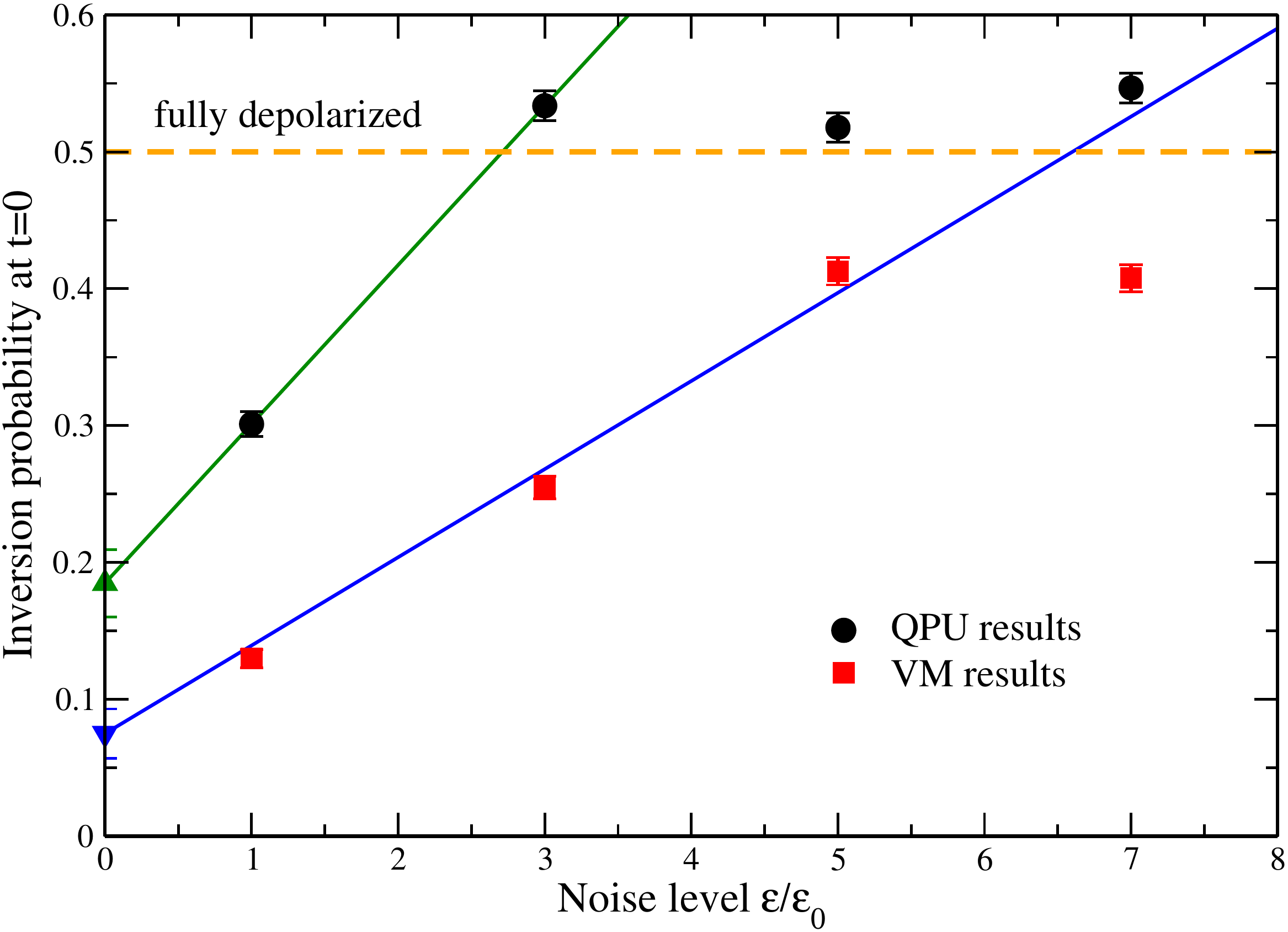}
 \caption{(Color online) Inversion probability $P_{\text{inv}}$ at the initial time $t=0$ for the first neutrino. Black solid circles are results from the Vigo QPU~\cite{IBMQ_Vigo} while the red squares correspond to results obtained using the VM with simulated noise. Also shown are extrapolations to the zero noise limit, for both the QPU (green line) and the VM (blue line), together with the extrapolated value (greed triangle up and blue triangle down respectively). The dashed orange line denotes the result for a maximally mixed state.}
\label{fig:pop_extrap}
\end{figure}

In order to highlight the difficulties encountered when performing noise extrapolation for this data, we plot in Fig.~\ref{fig:pop_extrap} the results obtained from both the QPU (black circles) and the VM (red squares) for the inversion probability of the first neutrino at the initial time $t=0$ together with a linear extrapolation using the first two points for the QPU (green line) and the first three points for the VM (blue line). The exact result is of course $P_{\text{inv}}(0)=0$ and we see that neither strategy is able to predict the correct value. The horizontal dashed line is the value expected when the system is in the maximally mixed state, corresponding to full depolarization. As shown in the data, for the real QPU results, only the first level of noise extrapolation contains useful information and a more gentle noise amplification strategy, like the one proposed in Ref.~\cite{He2020}, could provide a substantial advantage over the strategy adopted here.

\subsection{Dynamics of entanglement}

In order to track the evolution of entanglement in the system we perform complete state tomography for each of the $6$ possible qubit pairs in our system by estimating, for each pair $(k,q)$, the 16 expectation values
\begin{equation}
\label{eq:simple_mat_els}
M^{k,q}_{\alpha,\beta}(t) = \langle \Phi(t)\lvert P_k^\alpha\otimes P_q^\beta \rvert \Phi(t)\rangle\;,
\end{equation}
with $P_k=\{\mathbb{1},X,Y,Z\}$ the basis for $U(2)$ and $\ket{\Phi(t)}$ the state obtained from $\ket{\Phi_0}$ by applying the time-evolution operator as in Eq.~\eqref{eq:schroedinger-eq}. In principle, we might reconstruct the density matrix for the pair of qubits $(k,q)$ directly from these expectation values as
\begin{equation}
\label{eq:naive_dm}
{\bm \rho}^D_{kq}(t) = \sum_{\alpha=1}^4\sum_{\beta=1}^4M^{k,q}_{\alpha,\beta}(t)P_k^\alpha\otimes P_q^\beta\;.
\end{equation}
In practice however, we can only estimate the matrix elements $M^{k,q}_{\alpha,\beta}(t)$ to some finite additive precision, and the approximation in Eq.~\eqref{eq:naive_dm} is not guaranteed to be a physical density matrix (positive definite and with trace equal to 1). In this work we use the common approach (see eg.~\cite{Banaszek1999}) of performing a maximum-likelihood (ML) optimization, while enforcing the reconstructed density matrix ${\bm \rho}^{ML}_{kq}(t)$ to be physical. We note in passing that it is possible to devise operator basis that are more robust than the choice used in Eq.~\eqref{eq:simple_mat_els} (see eg.~\cite{Czartowski2020}) but we didn't explore this further in our work.

In order to propagate the effect of statistical errors into the final estimator for ${\bm \rho}^{ML}_{kq}(t)$, we use a resampling strategy similar to what was introduced in~\cite{Roggero_nptodg} but using a Bayesian approach to determine the empirical posterior distribution. We provide a detailed description of the adopted protocol in Appendix \ref{app:posterior_sampling}.

\subsubsection{Entanglement entropies}

As we mentioned in the introduction, one of the main differences between a mean field description and the full many-body description of the dynamics of the neutrino cloud is the absence of quantum correlations, or entanglement, in the former. Past work on the subject~\cite{Cervia:2019,Rrapaj2020} looked at the single spin entanglement entropy defined as
\begin{equation}
 S_k(t) = - {\rm Tr} \left[{\bm \rho}_k(t)\log_2\left({\bm \rho}_k(t)\right)\right]\;,
\end{equation}
with ${\bm \rho}_k(t)$ the reduced density matrix of the $k$-th spin. A value of the entropy $S_k(t)$ different from zero indicates the presence of entanglement between the $k$-th neutrino and the rest of the system.

In our setup, we compute the one-body reduced density matrix from the maximum-likelihood estimator of the pair density matrix defined above, explicitly
\begin{equation}
 S^{ML}_{k;q}(t) = - {\rm Tr} \left[{\bm \rho}^{ML}_{k;q}(t)\log_2\left({\bm \rho}^{ML}_{k;q}(t)\right)\right]\;,
\end{equation}
where the reduced density matrices are computed from
\begin{equation}
{\bm \rho}^{ML}_{k;q}(t) = {\rm Tr}_q \left[{\bm \rho}^{ML}_{kq}(t)\right]\;,
\end{equation}
and ${\rm Tr}_q$ denotes the trace over the states of the $q$-th qubit.
We combine the 3 values obtained in this way for each neutrinos as follows: the estimator for the single-spin entanglement entropy is obtained from the average 
\begin{equation}
S^{\text{avg}}_k(t) = \frac{1}{3} \sum_{q}S^{ML}_{k;q}(t)\;,
\end{equation}
summing over pairs containing the k-th spin, while as an error estimate we use the average of the 3 errors.

\begin{figure}
 \centering
 \includegraphics[width=0.49\textwidth]{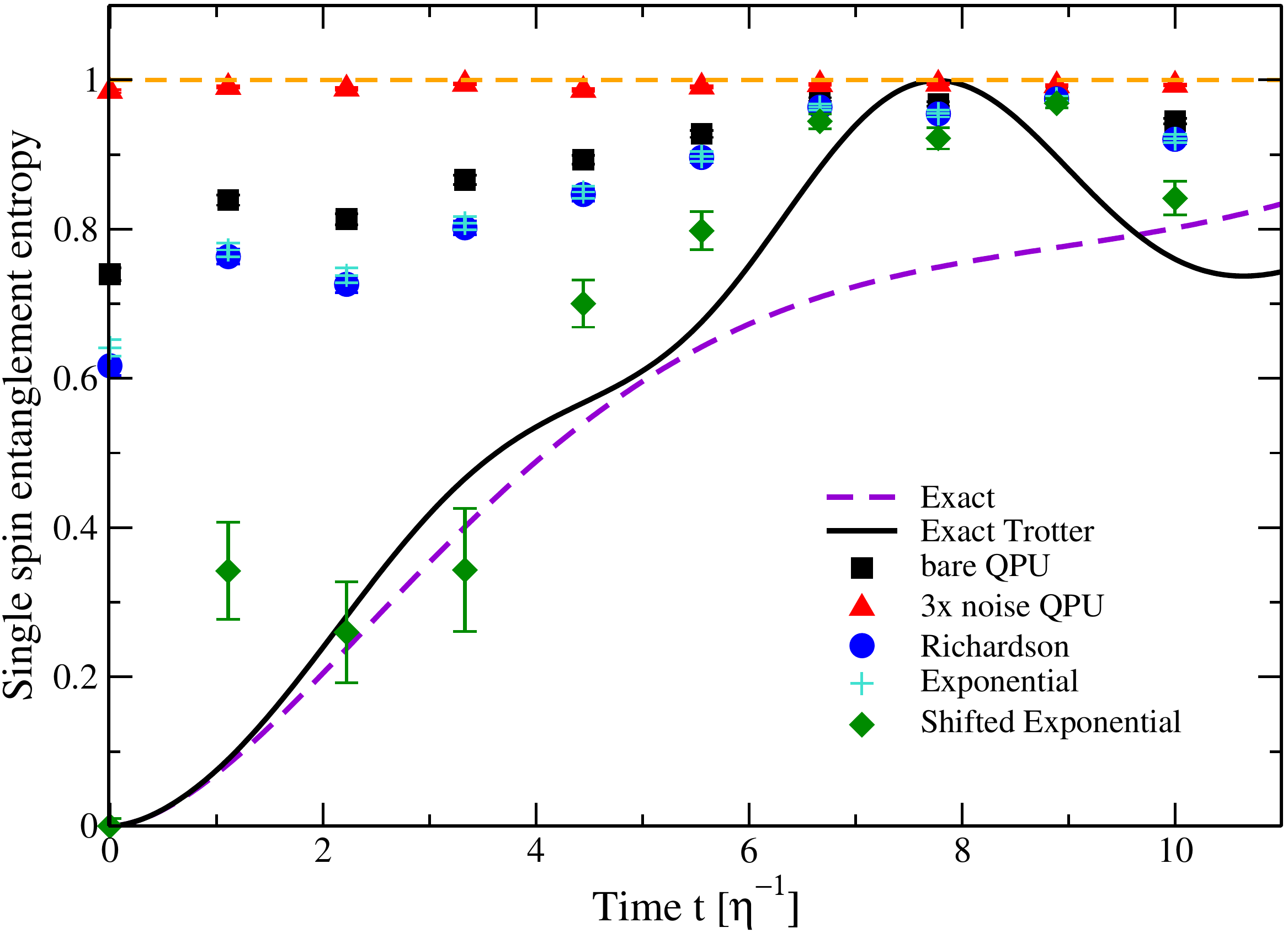}
 \caption{(Color online) Single spin entanglement entropy for neutrino 2. Black square are bare results obtained from the QPU, red triangles are results obtained by amplifying the noise to $\epsilon/\epsilon_0=3$, the blue circles are obtained using Richardson extrapolation, the turquoise plus symbols indicate results obtained using the standard exponential extrapolation and the green diamonds correspond to the results obtained from a shifted exponential extrapolation using the maximum value of the entropy (indicated as a dashed orange line).}
\label{fig:sq_ent1}
\end{figure}

As for the case of the inversion probability $P_{\text{inv}}(t)$ studied in the previous section, the substantial noise present in the QPU data prevents us from using the full set of results at the 4 effective noise levels. In order to overcome this difficulty, we have performed zero noise extrapolations using only results for effective noise levels $r=\epsilon/\epsilon_0=(1,3)$ and performed a Richardson extrapolation (in this case equivalent to a simple linear fit as done in Ref.~\cite{Dumitrescu2018}), a two point exponential extrapolation~\cite{Endo2018}, and an exponential extrapolation with shifted data. The latter technique consists in shifting the data for the entropy by $-1$ (it's maximum value) so that the result, in the limit of large noise, tends to 0 instead of $\log_2(2)=1$. We then shift back the result obtained after extrapolation. The exponential extrapolation method is well suited for situations where expectation values decay to zero as a function of the noise strength $\epsilon$, while maintaining a consistent sign, and this shift allows us to make the data conform to this ideal situation (see Appendix~\ref{app:error_mit} for more details on the method). The impact on the efficacy of the error mitigation is dramatic as can be seen in the results presented in Fig.~\ref{fig:sq_ent1} for the entropy of the second neutrino (the entropies for the other neutrinos follow a similar pattern; see Appendix~\ref{app:sq_ent} for all four results). The results with the standard exponential extrapolation are presented as the turquoise plus symbols, they are almost the same as those obtained using Richardson extrapolation (blue circles) and show a significant systematic error. On the contrary, the results obtained with the Shifted Exponential extrapolation (green diamonds) are much more close to the expected results with our pair propagator (solid black curve). We expect more general multi-exponential extrapolation schemes, like those proposed in Refs.~\cite{giurgicatiron2020,cai2020}, to enjoy a similar efficiency boost in the large noise limit achieved with deep circuits.

\begin{figure}
 \centering
 \includegraphics[width=0.49\textwidth]{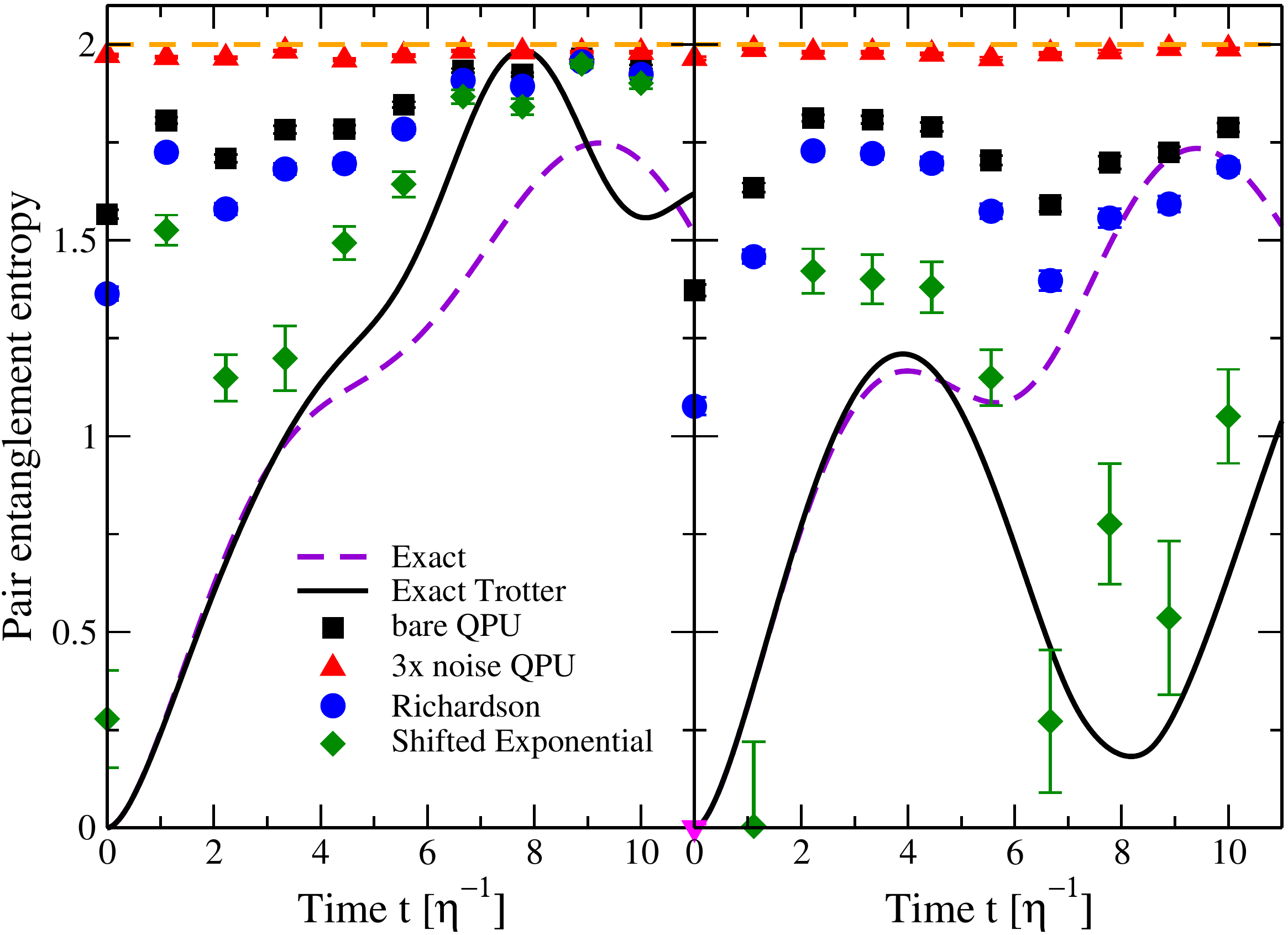}
 \caption{(Color online) Pair entanglement entropy for the neutrino pair $(1,2)$ starting as $\ket{e}\otimes\ket{e}$ (left panel) and pair $(2,4)$ which starts as the flavor state $\ket{e}\otimes\ket{x}$ (right panel). Results obtained directly from the QPU are shown as black squares ($r=1$) and red triangles ($r=3$) while blue circles and green diamonds indicate mitigated results using Richardson and the Shifted Exponential extrapolations respectively. For the Shifted Exponential ansatz we use the maximum value of the entropy (indicated as a dashed orange line).The magenta triangle indicates a mitigated result with Shifted Exponential extrapolation below zero within errorbars.}
\label{fig:pair_ent_04}
\end{figure}

Using the reconstructed pair density matrix ${\bm \rho}^{ML}_{kq}(t)$, we can clearly also evaluate directly the entanglement entropy of the pair
\begin{equation}
 S^{ML}_{kq}(t) = - {\rm Tr} \left[{\bm \rho}^{ML}_{kq}(t)\log_2\left({\bm \rho}^{ML}_{kq}(t)\right)\right]\;.
\end{equation}
In Fig.~\ref{fig:pair_ent_04} we show the result of this calculation for the pair $(1,2)$, which started as electron flavor at $t=0$, and the pair $(2,4)$ which started instead as heavy flavor states.

\subsubsection{Concurrence}

In order to better understand these quantum correlations, we also compute the concurrence~\cite{Wooters1998} for all the pair states. This measure of entanglement is defined for a 2 qubit density matrix as
\begin{equation}
\label{eq:concurr}
C(\rho) = \max\left\{0,\lambda_0-\lambda_1-\lambda_2-\lambda_3\right\}\;,
\end{equation}
where $\lambda_i$ are the square roots of the eigenvalues, in decreasing order, of the non-Hermitian matrix
\begin{equation}
M = \rho\left(Y\otimes Y\right)\rho^*\left(Y\otimes Y\right)\;,
\label{eq:concurr_M}
\end{equation}
with the star symbol indicating complex conjugation. The usefulness of this measure is its relation with the entanglement of formation~\cite{Hill1997,Wooters1998}, which is the minimum number of maximally-entangled pairs needed to represent $\rho$ with an ensemble of pure states~\cite{Hill1997}. 

\begin{figure}[tbh]
 \centering
 \includegraphics[width=0.49\textwidth]{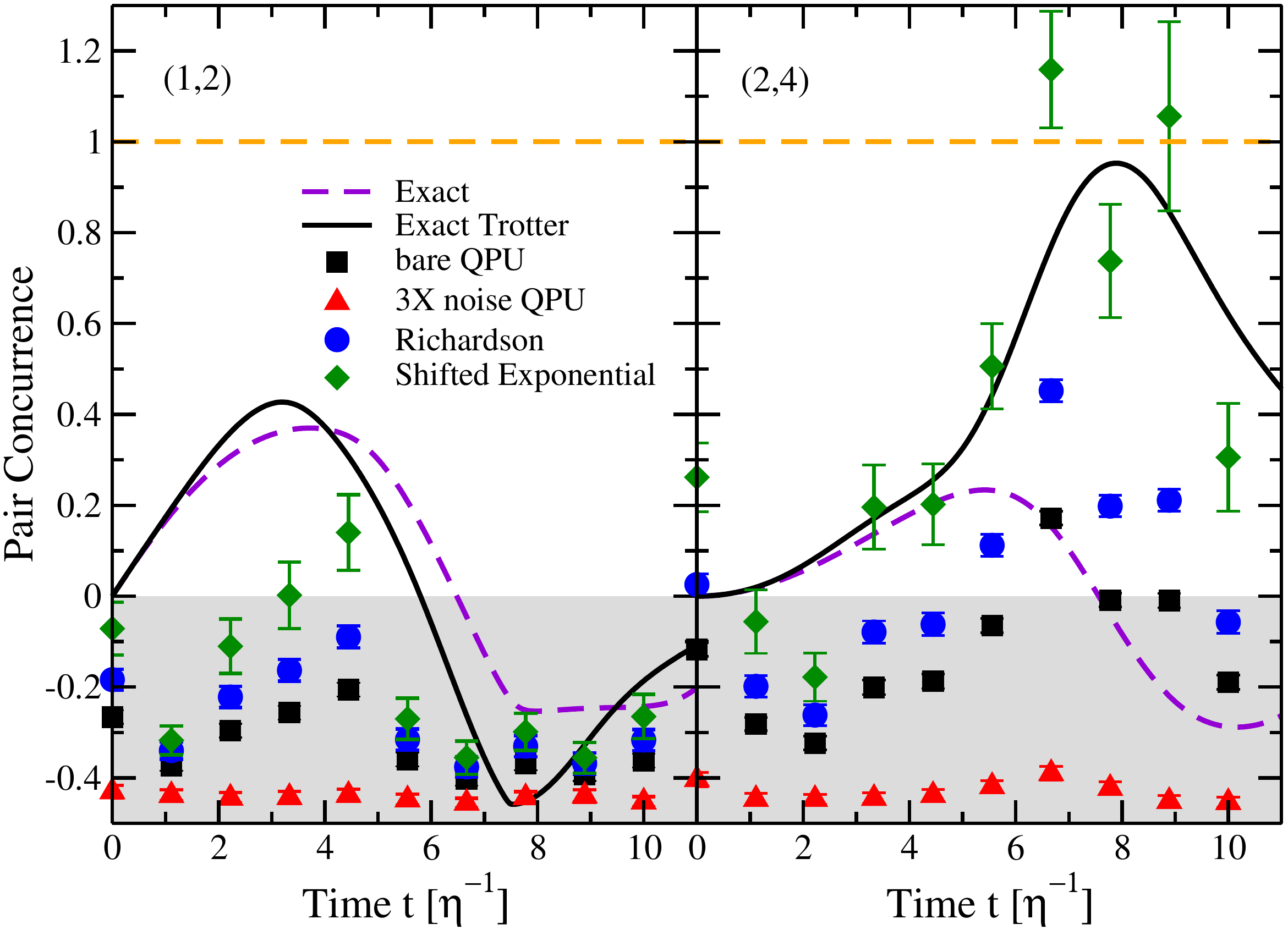}
 \caption{(Color online) Extended concurrence $\widetilde{C}$ for two pairs of neutrinos, $(1,2)$ in the left and $(2,4)$ in the right panel. The convention for the curves and date point used here is the same as in Fig.~\ref{fig:pair_ent_04}. The gray area indicates the region where the concurrence $C(\rho)$ is zero. The maximum value for the concurrence is shown as a dashed orange line.}
\label{fig:conc_pair_04}
\end{figure}

The definition of concurrence in Eq.~\eqref{eq:concurr} does not lend itself as easily to be adapted in an error extrapolation procedure as the one we used to obtain the mitigated results in the previous sections. This is due to the presence of the $max$ function in the definition of the concurrence: when the error is sufficiently strong to make the difference in eigenvalues
\begin{equation}
\widetilde{C}(\rho) = \lambda_0-\lambda_1-\lambda_2-\lambda_3
\end{equation}
negative, the concurrence in Eq.~\eqref{eq:concurr} ceases to carry information about the error free result. For this reason, we will regard $\widetilde{C}$ as an ``extended concurrence" which varies smoothly for large error levels and perform the truncation to positive values only after the zero noise extrapolation.
The results obtained from the simulation on the Vigo QPU are shown in Fig.~\ref{fig:conc_pair_04} for two pairs of neutrinos: pair $(1,2)$ starting as like spin at $t=0$ and pair $(2,4)$ which started as opposite flavors. The complete set of results for all pairs can be found in Fig.~\ref{fig:sq_all} in Appendix~\ref{app:sq_ent}.

The bare results are shown as black squares and we can immediately notice why the definition of $\widetilde{C}$ is so important in our case: the only bare data point with a measurable concurrence $C(\rho)$ is at $t\approx6.7\eta^{-1}$ for pair $(2,4)$ (the right panel in Fig.~\ref{fig:conc_pair_04}) while all the other results, including those obtained with a larger noise level (red triangles), are compatible with zero. In this situation, no mitigation of $C(\rho)$ would be possible.

By keeping the negative contributions, we see that the bare results often contain a substantial signal, while those at a higher error rate are already almost at the asymptotic value $\widetilde{C}=-0.5$ expected for a completely depolarized system~\footnote{Note that our results seem to converge to a larger asymptotic value of {$\widetilde{C}\approx-0.44$} instead of {$\widetilde{C}=-0.5$}. We can empirically explain this difference as the effect of statistical fluctuations.}. This allowed us to perform error extrapolation using both the Richardson and Shifted Exponential ansatz. Similarly to what we observed for the entanglement entropies in the previous section, the Shifted Exponential ansatz (with shift $-0.5$) produces consistently better results than Richardson extrapolation. This indicates that we are more close to the asymptotic large error regime than the small error limit used to motivate a polynomial expansion. The resilience of the exponential extrapolations to large errors, especially augmented by an appropriate shift, is seen here to be critical in extracting physical information from quantum simulations carried out near the coherence limit of the device used for the implementation.    

\section{Conclusions}
\label{sec:conclusion}

In this work, we presented the first digital quantum simulation of the flavor dynamics in collective neutrino oscillations using current quantum technology. The results reported for the evolution of flavor and entanglement properties of a system with $N=4$ neutrino amplitudes show that current quantum devices based on superconducting qubits are starting to become a viable option for studying out-of-equilibrium dynamics of interacting many-body systems. The reduced fidelity in the results obtained here, compared to the simulations reported previously in Ref.~\cite{Roggero_nptodg} employing the same quantum processor and a comparable number of entangling gates, points to the importance of controlling unitary errors associated with the imperfect implementation of arbitrary single-qubit rotations (on average $<1\%$ for the device used in both works). In future work we plan to explore the use of more advanced error mitigation strategies, such as Pauli twirling~\cite{Wallman2016} or symmetry protection~\cite{Tran2021}, to achieve a better overall fidelity.

We showed the zero-noise error extrapolation using a shifted Gaussian ansatz to be remarkably efficient in predicting the expected error-free estimator of observables. Given the large circuits employed in this work, past experience with zero-noise extrapolations (see eg.~\cite{roggero2020A,Roggero_nptodg}) suggest the exponential ansatz to be appropriate due to the large noise rates, and we find it to indeed outperforms Richardson extrapolation in this regime. The current results highlight the importance of using alternative measures of entanglement to the entropy in order to extract reliable information about quantum correlations in the states generated on the quantum device. Using the pair concurrence together with the entropy provides a robust way to detect entanglement even in the presence of substantial noise, like in the results shown here. We expect these insights, and the mapping of the neutrino evolution problem into a swap network, to prove very valuable in future explorations of out-of-equilibrium neutrino dynamics with near-term, noisy, quantum devices.

\begin{acknowledgments}
This work was supported by the InQubator for Quantum Simulation under U.S. DOE grant No. DE-SC0020970, by the Quantum Science Center (QSC), a National Quantum Information Science Research Center of the U.S. Department of Energy (DOE), by the U.S. Department of Energy under grant No. DE-FG02-00ER41132, DE-SC0021152 and by the U.S. National Science Foundation under Grants No. PHY-1404159 and PHY-2013047. 
Benjamin Hall acknowledges support from the U.S. Department of Energy (DOE) through a quantum computing program sponsored by the Los Alamos National Laboratory (LANL) Information Science \& Technology Institute.
We  acknowledge  use  of the  IBM  Q  for  this  work. The  views  expressed  are those of the authors and do not reflect the official policy or position of IBM or the IBM Q team. 
\end{acknowledgments}

\appendix
\section{Pair product propagator}
\label{app:pair_prop}

In this appendix, we provide additional details about the pair approximation of the time-evolution operator $U_2(t)$ introduced in Eq.~\eqref{eq:pair_prop_app} of the main text. In particular, we will present a direct comparison with the simpler approximation $U_1(t)$ corresponding to the canonical first order Trotter-Suzuki step.

As mentioned in the main text, the asymptotic scaling of the approximation error $\epsilon$ is quadratic in the time-step $t$ for both approximations~\cite{Suzuki91}. The pair approximation is expected, however, to perform better in practice for cases where an accurate description of pair evolution is important due, for instance, to strong cancellations between the one-body and two-body contributions in the Hamiltonian. In the neutrino case, these situations can occur with appropriate initial conditions so that, for typical states in the evolution, we have for most pairs
\begin{equation}
\label{eq:pairp_err}
\left|\langle K_{pq}\rangle+\langle V_{pq}\rangle\right|\ll\left|\langle K_{pq}\rangle\right|+\left|\langle V_{pq}\rangle\right|\;,
\end{equation}
where we used the short-hand (cf. Eq.~\eqref{eq:ham_decomp} in main text)
\begin{equation}
\langle K_{pq}\rangle = \frac{\vec{b}}{N-1}\cdot\langle\vec{\sigma}_k+\vec{\sigma}_q\rangle\quad\langle V_{pq}\rangle=J_{kq}\langle\vec{\sigma}_k\cdot\vec{\sigma}_q\rangle\;.
\end{equation}

\begin{figure}
 \centering
 \includegraphics[width=0.49\textwidth]{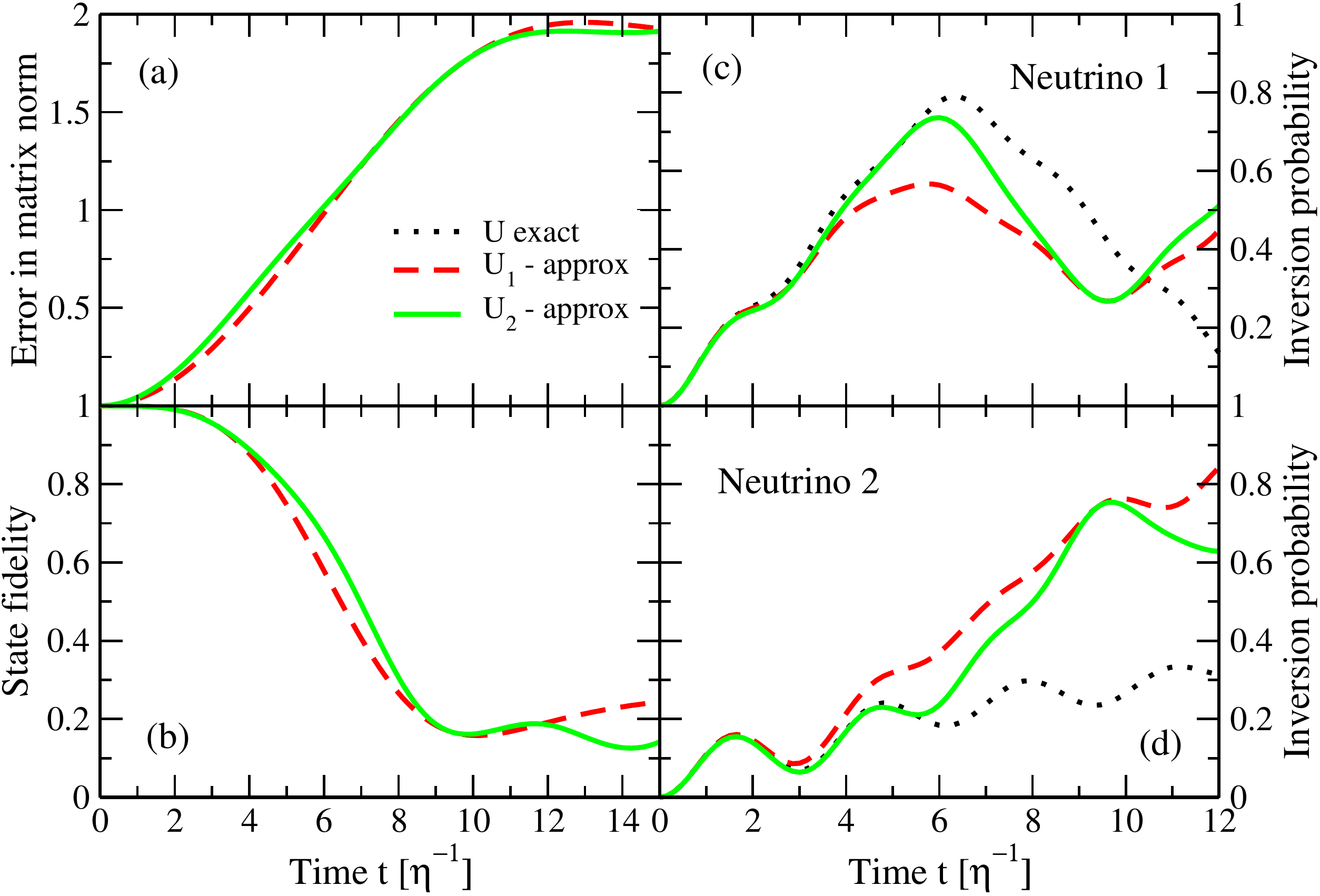}
 \caption{(Color online) Panel (a) shows the error in matrix 2-norm Eq.~\eqref{eq:mnorme} of the two approximations $U_1$ and $U_2$ described in the text. Panel (b) shows the state fidelity and the right panels show results for the inversion probability $P_{\text{inv}}(t)$. Panel (c) is for neutrino 1 while panel (d) is for neutrino 2.}
\label{fig:prop_cmp}
\end{figure}

Since the difference between the two approximation is not expected to hold for a generic initial state, standard error measures like the matrix norm of the difference with the exact propagator
\begin{equation}
\label{eq:mnorme}
\left\|\exp\left(-itH\right)-U_{1/2}(t)\right\|
\end{equation}
are not expected to capture the effect. This is in fact found in practice for our system. In panel (a) of Fig.~\ref{fig:prop_cmp} we show the estimate from Eq.~\eqref{eq:mnorme} for the $N=4$ neutrino model considered in this work. This error estimate indicates that the $U_1$ approximation has a smaller maximum error than $U_2$ up to long times. We can look at a more direct measure of accuracy for our specific setup by considering instead the state fidelity 
\begin{equation}
f(t) = \left|\langle \Psi_{1/2}(t)|\Psi(t)\rangle\right|^2   
\end{equation}
between the exact state $\ket{\Psi(t)}$ at time $t$ and one of its approximations $\ket{\Psi_{1/2}(t)}$ obtained using either $U_1$ or $U_2$. We show $f(t)$ for both approximations in panel (b) of Fig.~\ref{fig:prop_cmp}. The result here suggest that instead the pair approximation produces a state with a higher fidelity than the simple linear propagator $U_1$, especially at relatively long time-steps $t\in[4,8]$.

Finally, since we are mostly interested in flavor observables diagonal in the computational basis, we also show a direct comparison of the inversion probability for two out of the $N=4$ neutrinos using both approximations and the exact propagator (panels (c) and (d)). These results show more clearly that the pair approximation allows us to correctly describe the evolution of flavor for substantially longer times than the canonical $U_1$ approximation. The results reported here do depend on the specific choice of ordering of qubits in the time evolution layers shown in Fig.~\ref{fig:swap_network}. In both the present analysis and the simulation results in the main text we used the best ordering which we empirically found to be $(1,3,2,4)$ as one would've expected based on the initial state and the criterion Eq.~\eqref{eq:pairp_err} above. 

A more rigorous discussion of the relative accuracy between the canonical first order and the pair approximation, together with the effect of ordering choices, will be explored in future work.

\section{Error mitigation}
\label{app:error_mit}
In the following subsections we describe in more detail the error mitigation techniques used in this work. 

\subsection{Propagation of statistical uncertainties}
\label{app:posterior_sampling}

In this section we describe the procedure we have adopted for propagating statistical errors in the results reported in the main text. We found that careful treatment of statistical errors was important for non linear functions of the expectation values like entropy and concurrence of a reconstructed density matrix.

In the following, we will symbolically denote as $\langle O\rangle$, expectation values of Pauli operators which can be measured directly on the device. These are, for instance, the expectation values $\langle X X\rangle$, $\langle X Y\rangle$, etc. needed to reconstruct a two-qubit density matrix.

We use a Bayesian approach to perform inference from the bare counts obtained from the device. The idea is best described initially for the simple case of a single qubit measurement. The probability of obtaining $m$ measurements of the state $\ket{1}$ out of a total of $M$ trials can be modelled as a binomial distribution
\begin{equation}
P_b(m;p) =     \binom{M}{m}p^{m}(1-p)^{M-m}\;,
\end{equation}
with $p$ the probability of a $\ket{1}$ measurement. In order to infer the parameter $p$ from a given sample $m_i$ of measurement outcomes, we use Bayes theorem
\begin{equation}
P(p|m_i) = \frac{P(m_i|p)P(p)}{\int dq P(m_i|q)P(q)}\;.
\end{equation}
For the single qubit measurement, we use the binomial distribution as likelihood $P(m_i|p)$ and, in order to obtain a posterior $P(p|m_i)$ in closed form, we use the conjugate prior of the binomial: the beta distribution
\begin{equation}
P_\beta(p;\alpha,\beta) = \frac{\Gamma(\alpha+\beta)}{\Gamma(\alpha)\Gamma(\beta)} p^{\alpha-1}(1-p)^{\beta-1}\;.
\end{equation}
Here $\alpha,\beta>0$ are the parameters defining the distribution and with $\alpha=\beta=1$ we obtain a uniform distribution. The advantage of using the Beta distribution as a prior is that, after a measurement $m_i$ of the system is available, the parameters $(\alpha_0,\beta_0)$ of the prior distribution get updated as
\begin{equation}
\alpha_i = \alpha_0 + m_i\quad\beta_i=\beta_0 + M-m_i\;.
\end{equation}
Intuitively we can interpret the parameters $(\alpha_0,\beta_0)$ of the prior as assigning an a-priori number of measurements to the measurement outcomes, which are then updated as more measurements are performed. In this work we used a simple uniform prior corresponding to the choice $\alpha_0=\beta_0=1$ for the prior parameters.

After the inference step described above, we calculate the expectation value of a generic non-linear function $\langle F[O]\rangle$ by sampling new outcomes $m'_k$ using the posterior distribution. More in detail, we generate a new artificial measurement $m'_k$ after the measured $m_i$ by the following procedure
\begin{itemize}
    \item sample a value $p'_k$ from the posterior $P(p'_k\lvert m_i)$ 
	\item sample a new measurement outcome $m_k'$ from the likelihood $P_b(m_k^\prime;p'_k)$
\end{itemize}

The new measurements $m_k'$ obtained in this way are then samples from the predictive posterior distribution.

Using an ensemble of size $L$ obtained in this way, we compute $\langle F[O]\rangle$ by taking an average of the results obtained for each individual sample
\begin{equation}
\langle F[O]\rangle \approx \frac{1}{L} \sum_{k=1}^L F[O_k]\;.
\end{equation}
The error bars reported in the main text are $68\%$ confidence intervals which we found in most cases where well approximated by a Gaussian approximation.

This scheme is complete only for single qubit measurements but a generalization to generic multiqubit observables can be obtained in a straightforward way. In the situation where we are estimating expectation values over $N$ qubits, the probability of measuring  a specific collection of $N$ bit strings $m_i$ in $M$ repeated trials can be described with a multinomial distribution with $N$ probabilities. We use this distribution as the likelihood $P(m_i|\vec{p})$ in Bayes theorem and, for similar reasons as above, we take its conjugate prior distribution: the Dirichlet distribution (also initialized as uniform as for the Beta above). The procedure we follow is otherwise exactly equivalent to what we described above.

\subsection{Read-out mitigation}
The qubit measurements on a real device are not perfect and it is therefore important to understand the associated systematic errors. We refer the reader to Appendix.~H.1 of Ref.~\cite{Roggero_nptodg} for a more detailed derivation of the exact procedure we employ and the motivations behind it. Here, we instead describe the main difference with the scheme described there which comes from the use of the Bayesian inference scheme described in the previous subsection.

In the calculations presented here, we work under the assumption that read-out errors are independent on each qubit and perform a set of $2N$ calibration measurements $c_i$ (requiring two separate executions) to extract the parameters $(\vec{e}_0,\vec{e}_1)$ of the noise model (see Eq.(H1) of Ref.~\cite{Roggero_nptodg}). In order to consistently propagate the statistical uncertainties associated from the finite sample statistic used to estimate the noise parameters, we use an additional layer of Bayesian sampling using a binomial prior for the two error probabilities $(e^n_0,e^n_1)$ associated to each qubit $n$. 

Using a single pair of error probability vectors $\epsilon_i=(\vec{e}_0,\vec{e}_1)_i$, obtained either by direct measurement or by sampling from the posterior, we can generate a linear transformation $\mathbf{C}_i$ that maps a set of (in general multi-qubit) measurements $m_i$ to a new set $\widetilde{m}_i$ with reduced read-out errors (see Ref.~\cite{Roggero_nptodg} for more details).

The complete procedure that we use to generate an ensemble of measurements $\{\widetilde{m}'_i\}$ with read-out mitigation starting from a single calibration measurement $c_i$ and Pauli operator measurement $m_i$ is as follows
\begin{itemize}
    \item sample a value $p'_k$ from the posterior $P(p'_k\lvert m_i)$ 
	\item sample a new measurement outcome $m_k'$ from the likelihood $P_b(m_i^\prime;p'_k)$
	\item for each qubit $n=\{1,\dots,N\}$
	\begin{itemize}
	    \item sample a pair $(e^{\prime n}_0,e^{\prime n}_1)$ of error probabilities from the posterior $P(e^n_0,e^n_1|c_i)$ 
	\end{itemize}
	\item use the sampled error probabilities $(\vec{e}'_0,\vec{e}'_1)$ to generate the linear transformation $\mathbf{C}'_k$ 
	\item apply the sampled correction matrix $\mathbf{C}'_l$ to $m_k'$ to obtain the read-out mitigated estimator $\widetilde{m}'_k$
\end{itemize}

The resulting ensemble of measurements can be used directly to estimate expectation values and confidence intervals as described above. In this way, we avoid having to explicitly construct the variance of the correction matrix $\mathbf{C}'_l$ using maximum likelihood estimation and then propagating the error perturbatively to arbitrary observables as done in Ref.~\cite{Roggero_nptodg}.

\subsection{Zero-noise-extrapolation}
For observables like the inversion probability, we adopt the procedure developed in Ref.~\cite{Roggero_nptodg}. For entanglement observables we adopt a two point shifted exponential extrapolation that we briefly describe here. We denote the entanglement observable as $\langle F[O]\rangle^{(L)}(r)$ where $L$ is the number of samples used and $r$ denotes the noise level of the circuit, proportional to the number of CNOT gates in the circuit. We first note that in the case of very high noise levels, denoted here with $\langle F[O]\rangle(r\rightarrow\infty)$ the density matrix corresponds to the maximally mixed state given by $\mathbb{1}/4$. Therefore, the concurrence in this case is $-1/2$ and the pair entanglement saturates to $2$.

Using an estimate for the large noise expected value $\langle F[O]\rangle(r\rightarrow\infty)$, we can then consider a simple exponential extrapolation of the form
\begin{equation}
\langle F[O]\rangle^{(L)}(r)-\langle F[O]\rangle(r\rightarrow\infty) = A^{(L)}_F e^{-\alpha r}\, ,
\label{eq:exp-reshifted}
\end{equation}
with $\alpha $ and $A^{(L)}_F$ the parameters of the model which can be obtain using results at two different noise levels $r$ and $r^\prime$.
The zero-noise extrapolated result in this model corresponds to the limit $r\to0$ and is given simply by the estimated $A^{(L)}_F$. More explicitly this becomes
\begin{eqnarray}
A^{(L)}_F=\langle F[O]\rangle^{(L)}(r)\left(\frac{\langle F[O]\rangle^{(L)}(r^\prime)}{\langle F[O]\rangle^{(L)}(r)}\right)^{r/(r-r^\prime)}\, ,
\end{eqnarray}
and the zero noise extrapolated observable is
\begin{eqnarray}
\langle F[O]\rangle^{(L)}(0)=A^{(L)}_F+ \langle F[O]\rangle(r\rightarrow\infty)\, .
\end{eqnarray}
Finally, the estimated statistical error is obtained by calculating the standard deviation of the $L$ copies as above. 

\section{Additional data for concurrence and entanglement entropy}
\label{app:sq_ent}
Here we show the full set of results for both entanglement entropy and concurrence for all the other pairs of qubits not shown in the main text. We denote with a magenta triangle, data-points that fall below zero for the entropy as in Fig.~\ref{fig:conc_pair_04} of the main text.

\begin{figure}[h!]
 \centering
 \includegraphics[width=0.49\textwidth]{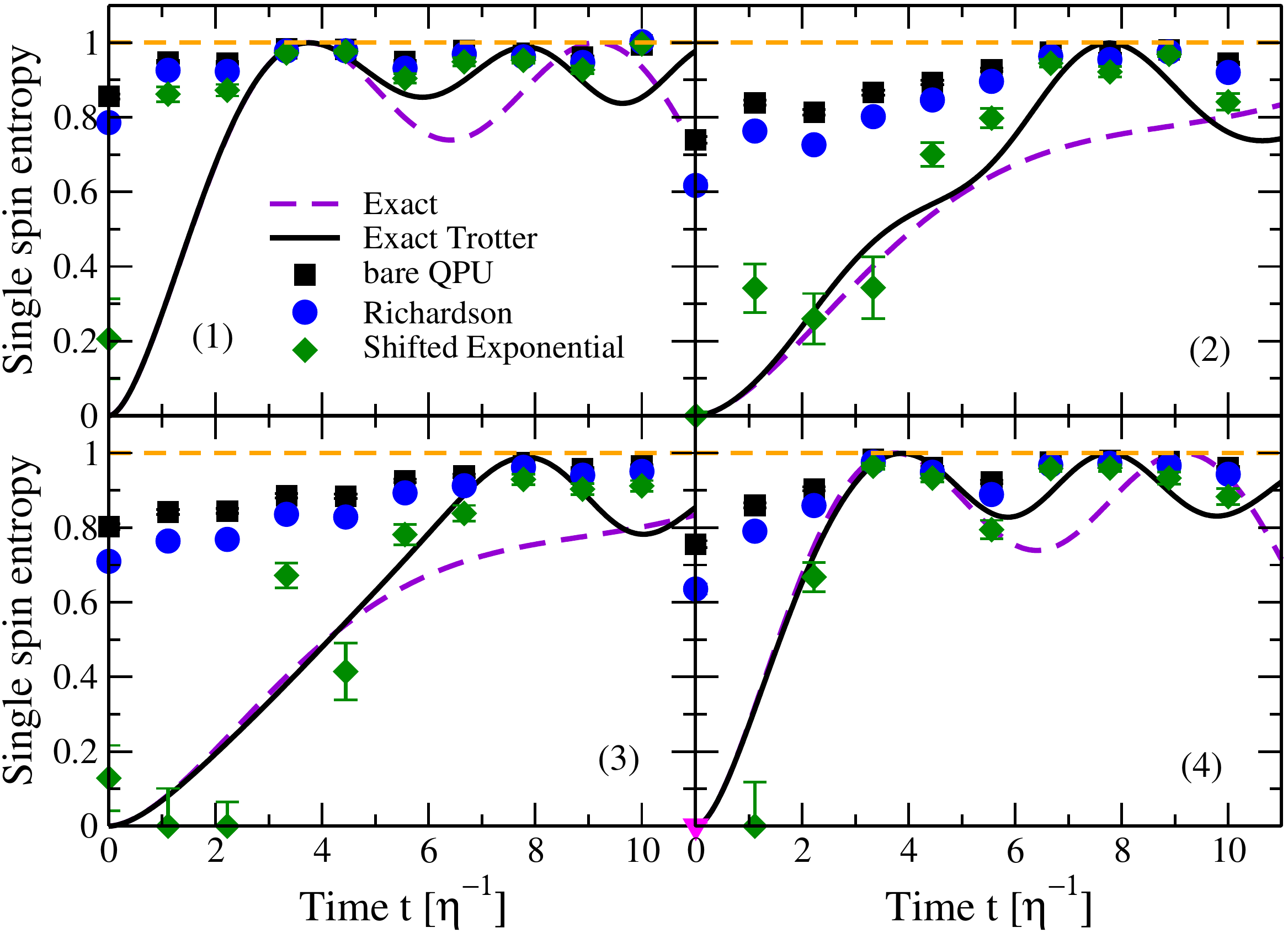}
 \caption{(Color online) Single spin entanglement entropy for all four neutrinos. Black square are bare results obtained from the QPU, the blue circles are obtained using Richardson extrapolation and the green diamonds correspond to the results obtained from a shifted exponential extrapolation using the maximum value of the entropy (dashed orange line). The magenta triangle indicates a mitigated result with Shifted Exponential extrapolation below zero within errorbars.}
\label{fig:sq_all}
\end{figure}

\begin{figure}[H]
 \centering
 \includegraphics[width=0.49\textwidth]{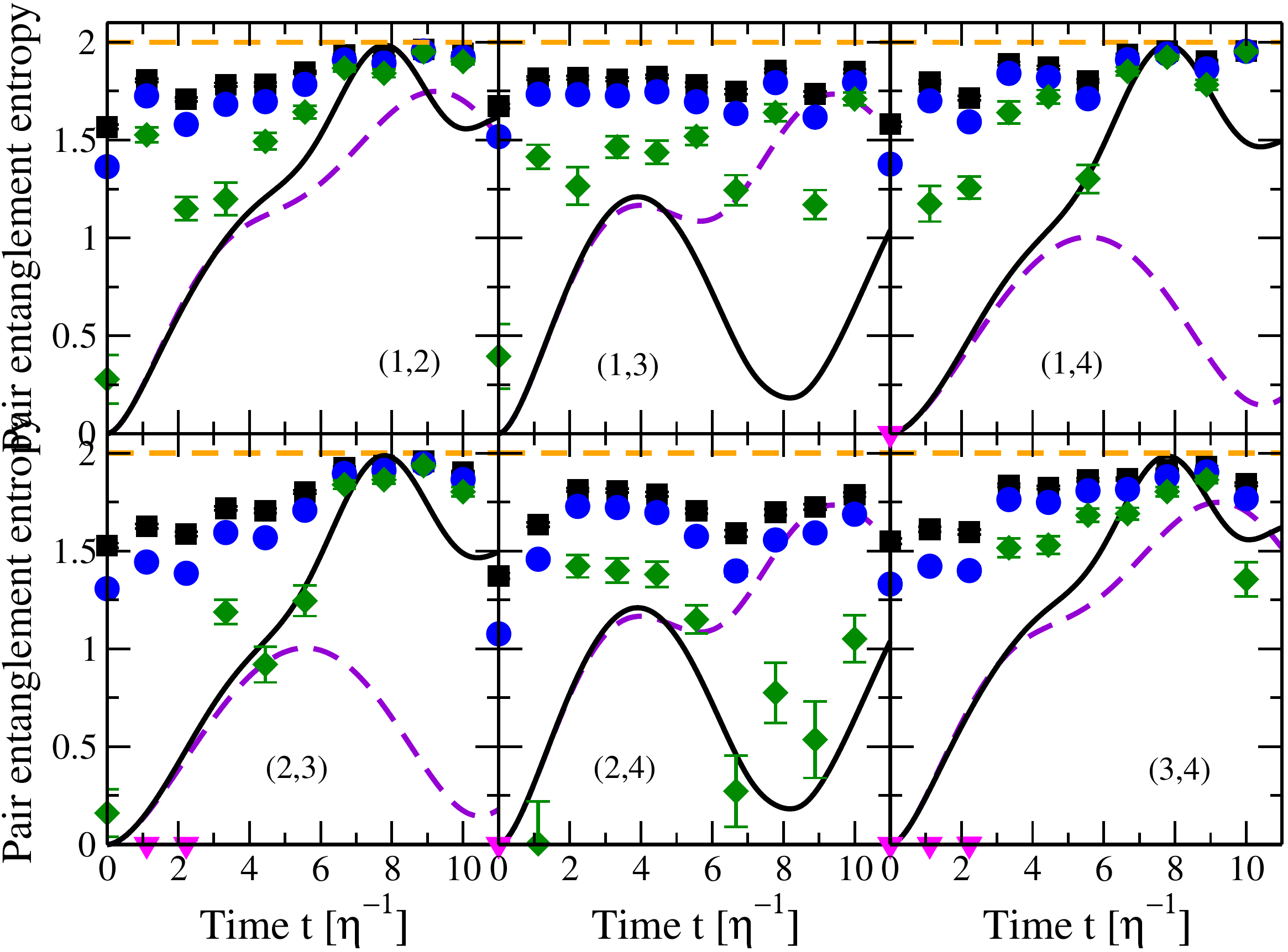}
 \caption{(Color online) Pair entanglement entropy for all pair of neutrinos. Black square are bare results obtained from the QPU, red triangles are results obtained by amplifying the noise to $\epsilon/\epsilon_0=3$, the blue circles are obtained using Richardson extrapolation and the green diamonds correspond to the results obtained from a shifted exponential extrapolation using the maximum value of the entropy (indicated as a dashed orange line). The magenta triangle points are mitigated results with Shifted Exponential extrapolation below zero within errorbars.}
\label{fig:pair_ent_all}
\end{figure}

\begin{figure}[H]
 \centering
 \includegraphics[width=0.49\textwidth]{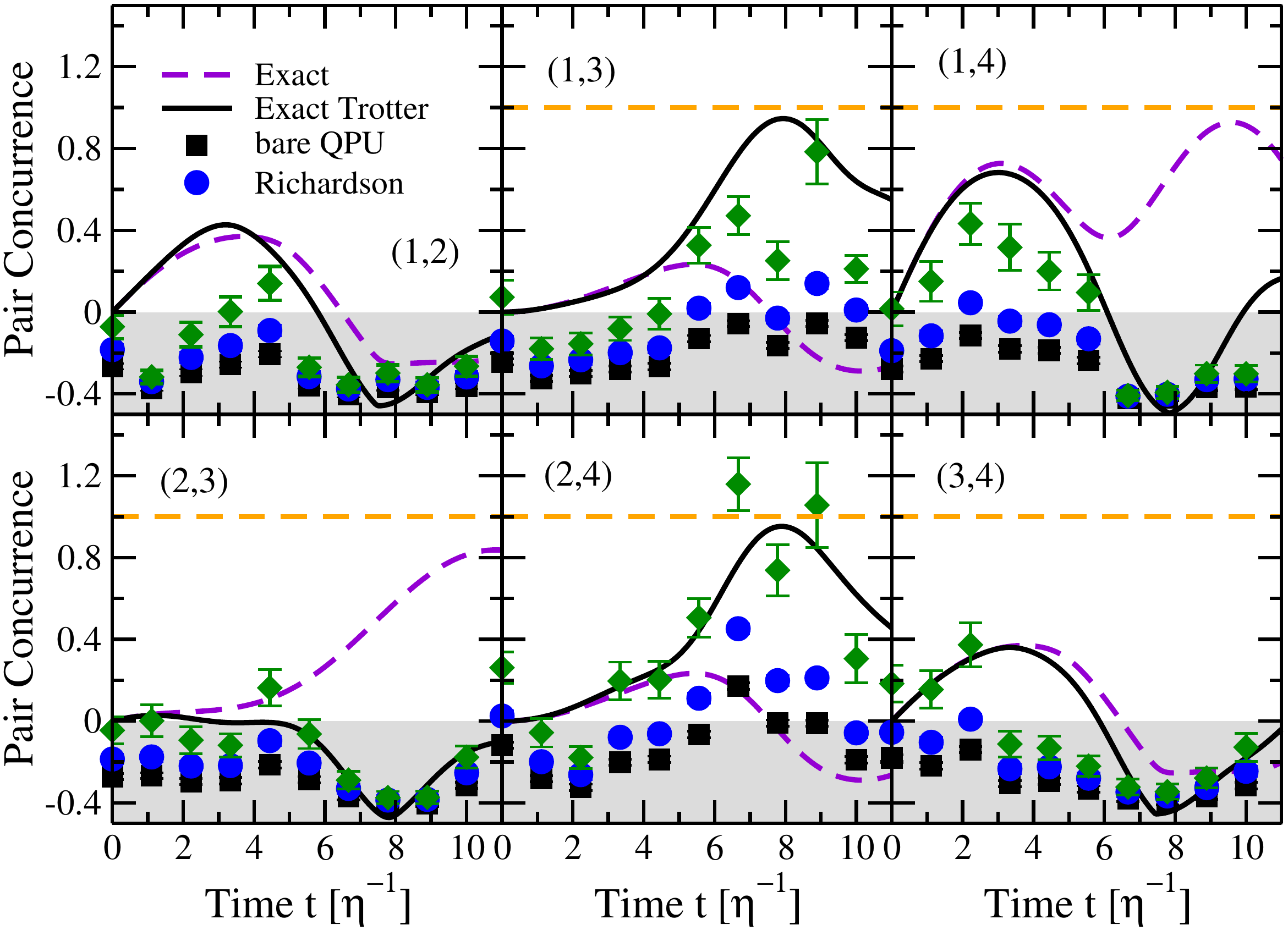}
 \caption{(Color online) Entanglement concurrence for all the pairs of qubits. The maximum value for the concurrence is shown as a dashed orange line.}
\label{fig:conc_pair_1}
\end{figure}

\end{document}